\documentclass[
amsmath,amssymb,showkeys]{elsarticle}
\usepackage{graphicx}
\usepackage{mathptmx}
\usepackage{epsfig}
\usepackage{units}
\usepackage{url}
\usepackage{upgreek}
\usepackage{wasysym}
\usepackage{hyperref}
\usepackage{lineno}

\begin{document}
\linenumbers

\begin{frontmatter}

\title{A new fission-fragment detector to complement the CACTUS-SiRi setup at the Oslo Cyclotron Laboratory}

\author[uio,atomki]{T.G.~Tornyi\corref{cor}} \ead{tornyitom@atomki.hu}
\author[uio]{A.~G\"{o}rgen}
\author[uio]{M.~Guttormsen}
\author[uio]{A.C.~Larsen}
\author[uio]{S.~Siem}
\author[atomki]{A.~Krasznahorkay}
\author[atomki,mpi]{L.~Csige}

\cortext[cor]{Corresponding author.}

\address[uio]{Department of Physics, University of Oslo, Norway}
\address[atomki]{Institute of Nuclear Research of the Hungarian Academy of Sciences (MTA Atomki), Debrecen, Hungary}
\address[mpi]{Max-Planck-Institute for Quantum Optics, D-85748 Garching, Germany}

\begin{abstract}
An array of Parallel Plate Avalanche Counters (PPAC) for the detection of heavy ions
has been developed. 
The new device, NIFF (Nuclear Instrument for Fission Fragments), 
consists of four individual detectors and covers $\unit 60{\%}$ of 2$\pi$. 
It was designed to be used in conjunction with the SiRi array of ${\Delta}E-E$ silicon telescopes
for light charged particles and fits into the CACTUS array of 28 large-volume NaI 
scintillation detectors at the Oslo Cyclotron Laboratory. 
The low-pressure gas-filled PPACs are sensitive for the detection of fission fragments, but are 
insensitive to scattered beam particles of light ions or light-ion ejectiles. 
The PPAC detectors of NIFF have good time resolution and can be used either to select or to 
veto fission events in in-beam experiments with light-ion beams and actinide targets. 
The powerful combination of SiRi, CACTUS, and NIFF provides new research opportunities for the 
study of nuclear structure and nuclear reactions in the actinide region. 
The new setup is particularly well suited to study the competition of fission and $\gamma$ 
decay as a function of excitation energy. 
\end{abstract}

\begin{keyword}
    PPAC \sep fission fragment detector \sep coincidence
\PACS 29.40.-h
\end{keyword}

\end{frontmatter}

\section{Introduction}

In-beam spectroscopy of heavy nuclei often requires the detection of fission fragments, 
either because the fission process itself or the fission fragments are to be investigated, 
or to study alternative decay processes where fission events need to be suppressed as 
unwanted background. 
Cross sections for nuclear reactions induced by neutrons or light charged particles on 
actinide nuclei are important for nuclear energy applications and their measurements 
usually require the detection of fission fragments. 
It is often difficult or even impossible to directly measure cross sections of 
neutron-induced reactions on short-lived actinides because the radioactivity of the target
sample would be prohibitively large. 
A surrogate method using charged-particle induced reactions to produce the same 
excited compound nucleus from a longer-lived target nucleus was originally proposed by 
Cramer and Britt \cite{cramer} and is now regularly used \cite{escher} to study the decay 
of the compound nucleus, which is thought to be independent of its production mechanism.

The Oslo Cyclotron Laboratory (OCL) at the University of Oslo hosts the highly efficient 
CACTUS array \cite{cactus} of large-volume NaI scintillation detectors coupled to the SiRi 
array \cite{siri} of silicon ${\Delta}E-E$ detector telescopes. 
CACTUS consists of 28 $5^"\!\times 5^"$ NaI detectors with a full-energy peak efficiency of 
$\unit 15{\%}$ at $\unit[1.3] {MeV}$. 
Each scintillator is placed at a distance of $\unit[22] {cm}$ from the target and mounted 
with a $\unit[10] {cm}$ thick conical lead collimator with an opening of 
$\diameter = \unit[70] {mm}$ at the front surface. 
The CACTUS array can be complemented by 
high-purity germanium detectors and large-volume LaBr$_3$(Ce) scintillator detectors.

The SiRi particle telescope system \cite{siri} comprises eight trapezoidal modules arranged 
in a lampshade geometry facing the target at a distance of $\unit[5] {cm}$ at an angle of 
45$^{\circ}$. 
Each telescope module consists of a $\unit[130] {\upmu m}$ thick silicon detector in the 
front and a $\unit[1550] {\upmu m}$ thick back detector allowing particle identification 
via ${\Delta}E-E$ measurements. 
The front detectors are segmented into eight curved strips, which allows measuring the 
scattering angle of the light-ion projectile with a resolution of $\unit[2]{^{\circ}}$. 
SiRi can either be mounted in forward direction covering scattering angles $\theta$ between 
$40$ and $\unit[54]{^{\circ}}$, or in backward direction covering angles between $126$ and 
$\unit[140]{^{\circ}}$. 
In experiments using direct reactions with proton, deuteron, $^{3}$He, or $^{4}$He beams 
from the Oslo Cyclotron, the detection of the charged ejectiles in SiRi provides a measure 
of the excitation energy of the binary reaction partner with a typical energy resolution of 
$\unit[150] {keV}$.

The combination of the SiRi-CACTUS setup with an efficient fission-fragment detector array is 
well suited for cross section measurements using the surrogate technique. 
A compact fission detector inside the CACTUS array also offers the possibility to perform other 
types of experiments where either a tag on fission events or a veto is required. 
In addition to fitting into the CACTUS array, it was a requirement that the new fission fragment
detector can be used together with the SiRi charged-particle telescopes, which limited the 
angular coverage to only one hemisphere. 
For this purpose a fission fragment detector based on Parallel Plate Avalanche Counters (PPAC) was 
built. 
In this paper we present the design and discuss the performance of the new detector.

\section{Design parameters}
\label{sec:design}

The most important design parameter for the new detector was a high detection efficiency for 
fission fragments. 
Energy and position resolution are not important for the purpose of a pure tagging or veto detector.
To allow the measurement of triple coincidences between light charged particles, $\gamma$-rays, 
and fission fragments, the fission fragment detectors should be fast with a time resolution of the 
order of nanoseconds. 
Ideally the detector should only trigger on heavy ions and be insensitive to light ions, electrons, 
and $\gamma$-rays. 
Given these requirements we have chosen to base the new detector on low-pressure gas filled PPAC 
\cite{stelzer,hempel,gaukler,breskin,ganz,sanabria}. 
It is a further advantage that PPAC detectors do not show ageing effects due to the continuous 
exchange of the gas, contrary to silicon detectors, for example, which would rapidly deteriorate 
with heavy-ion implantation. 
A fission-fragment detector based on PPAC detectors requires only small amounts of material inside 
the vacuum chamber, having no significant influence on the performance of the CACTUS detectors due 
to scattering of $\gamma$-rays. 

\begin{figure}
\begin{center}
\includegraphics[width=\linewidth]{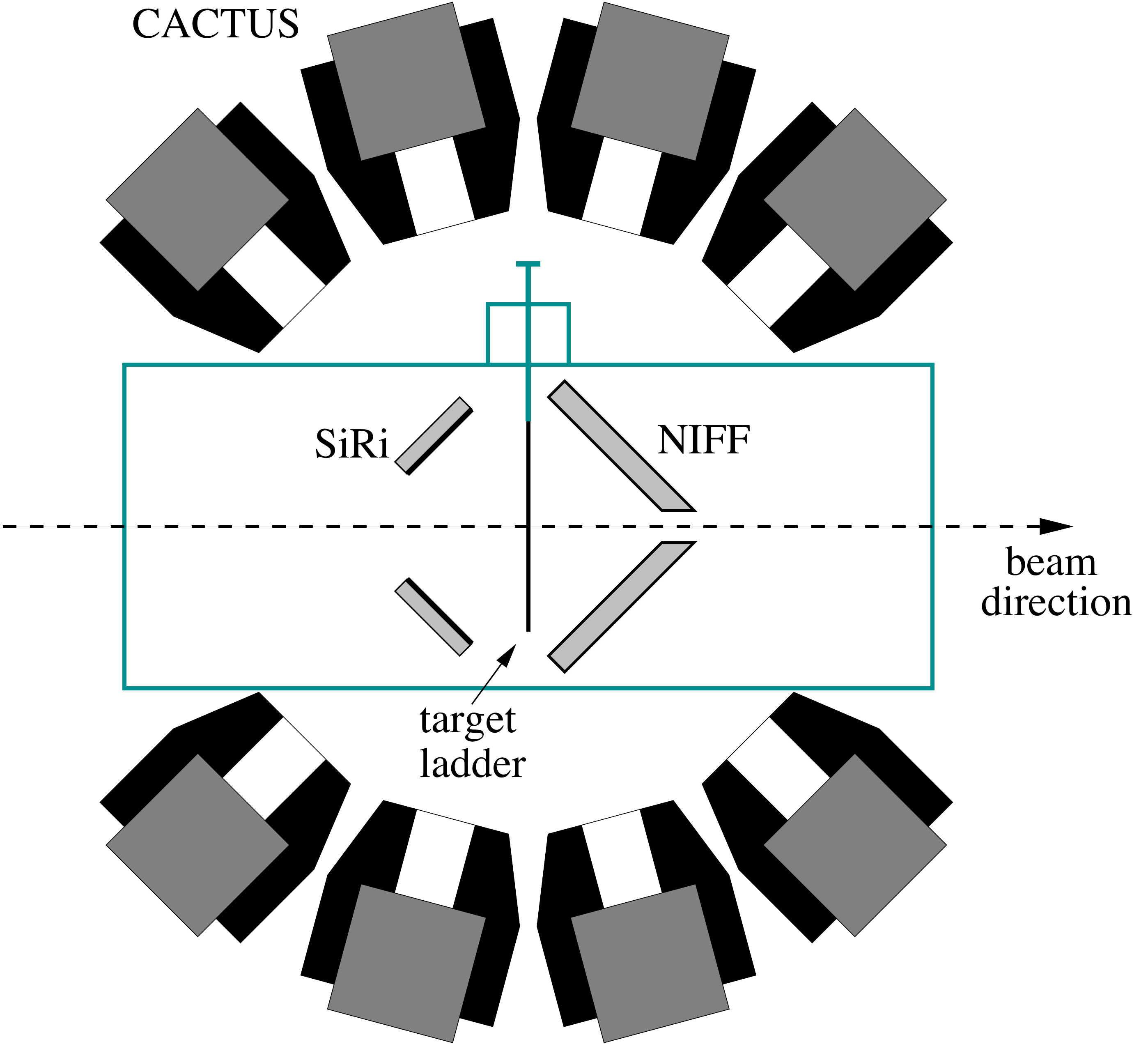}
\caption{Schematic view (not to scale) of the setup comprising the CACTUS array of 28 collimated NaI detectors surrounding the vacuum chamber with the SiRi particle telescope array mounted under backward angles and the new NIFF fission-fragment detector under forward angles. Targets are  inserted into the space between SiRi and NIFF via a target ladder.}
\label{fig:set_up}
\end{center}
\end{figure}

The design parameters of the PPAC detectors were strongly limited by the geometry and the size of the 
existing setup comprising CACTUS and SiRi. 
The geometry of the setup is shown schematically in Fig.~\ref{fig:set_up}. 
Since the silicon telescopes of SiRi together with their support structure cover either the forward or 
backward hemisphere with respect to the beam direction, only one hemisphere is available for the 
fission-fragment detector. 
Since the aim of the detector is to determine whether fission occurred or not, it is sufficient to 
detect only one of the two fission fragments, which are emitted in opposite direction in light-ion 
induced reactions. 
In this way it is possible to achieve high efficiency although less than $\unit 50{\%}$ of the solid 
angle is covered.

The NaI detectors of CACTUS are mounted at fixed positions and their collimators leave a cylindrical 
space for the target chamber, which has an inner diameter of $\unit[11.7] {cm}$ and a length of $\unit[48.0] {cm}$. 
A new chamber was designed that allows inserting a target ladder from the side into the space between SiRi 
and the fission detector, replacing the previously used rotating target changing system. 
The PPAC detectors were designed to cover the largest possible fraction of the forward hemisphere. 
However, an opening is needed to allow the beam to exit, and it is not necessary to cover angles close 
to $\unit[90]{^{\circ}}$ with respect to the beam axis, since the fission fragments that are emitted in 
this direction loose all their energy in the target or the target frame. 
The NIFF detector consists of four PPAC modules which are arranged like the leaves of a four-leaf clover, 
as illustrated in Fig.~\ref{fig:geom}. 
Each module has an overall length of $\unit[62.5] {mm}$ and an overall width of $\unit[77] {mm}$, where 
the outer part forms a sector of a circle with $\unit[44.5] {mm}$ radius. 
The modules are placed at an angle of $\unit[45]{^\circ}$ with respect to the beam axis. 
A square of $\unit[20] {mm}$ side length in the center of the detector allows the beam to exit. 
The active area of NIFF covers $\unit[\sim] 60{\%}$ of the forward hemisphere. 
Measurements with a $^{252}$Cf source and the performance of the detector are discussed in Sect.~\ref{sec:performance}.

\section{Detector Layout}
\label{sec:layout}

\begin{figure}
\begin{center}
\includegraphics[width=\linewidth]{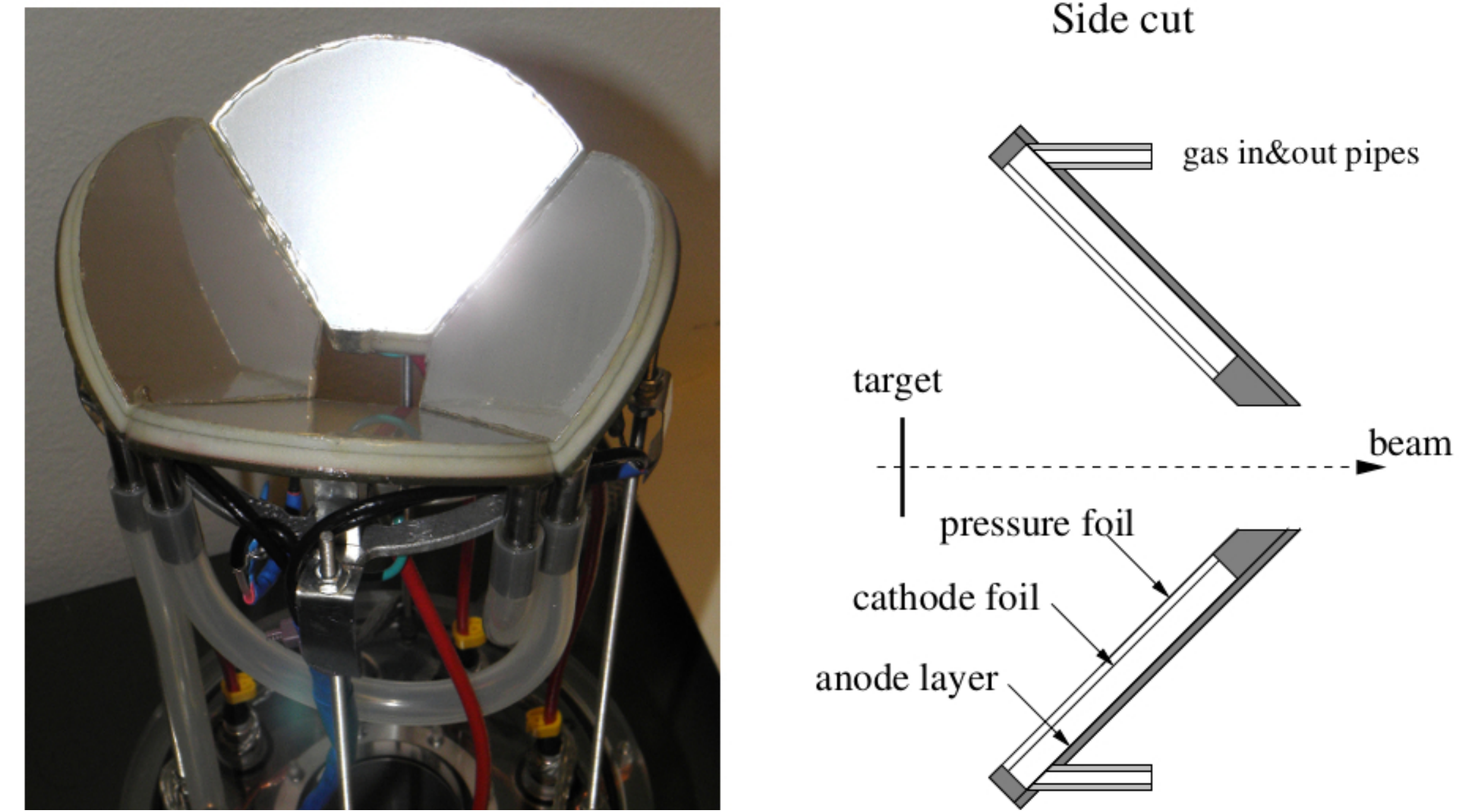}
\caption{Photo and cross section of NIFF detector. The dimensions of the PPAC modules and their arrangement are described in the text.}
\label{fig:geom}
\end{center}
\end{figure}

Each PPAC module consists of three main parts: an entrance window, the cathode foil, and the anode plate. 
The layout of the PPAC modules is shown in Fig.~\ref{fig:geom}. 
Both the entrance window and the cathode foil are made of $\unit[1.5] {\upmu m}$ thick Mylar foil, 
which is aluminised on one side. 
The thin aluminium layer is necessary to avoid the build-up of charge on the foils, which are coupled
to ground potential (see Figure~\ref{fig:electr}). 
The backplate of the PPAC modules is made from a single-sided printed-circuit board, with the polished 
copper layer of the sheet acting as anode of the PPAC. 
The modules are filled with high-purity ($> \unit 99.95{\%}$) isobutan (C$_4$H$_{10}$) gas at low 
pressure as the active material of the detector. 
The anode back-plate of each PPAC module has two holes of $\unit[2] {mm}$ diameter fitted with a small 
copper tube through which the gas enters and exits. 
The four modules are chained together with silicone tubing. In this way the gas is flowing subsequently 
through all four modules. 
The cathode foil is held by a PVC frame at a distance of $\unit[3.5] {mm}$ from the anode plate.

The pressure difference between the gas inside the PPAC module and the vacuum of the target chamber 
would cause the cathode foil to bulge outward. 
To obtain a uniform efficiency the cathode foil must be parallel with the anode back plate. 
To achieve this, a second, identical foil is placed on top of the cathode foil at a distance of 
$\unit[2] {mm}$ using a second PVC frame. 
The volumes between the two Mylar foils and between the cathode foil and anode back plate are connected
via a small hole in the frame. 
In this way only the outer foil bulges outward, while the cathode foil remains parallel with the anode. 
The frames that hold the Mylar foils in place have a width of $\unit[3] {mm}$, forming a $\unit[3] {mm}$
wide rim around each module that is insensitive to fission fragments. 
The total active area for each PPAC module is approximately $\unit[2240] {mm^2}$. 

\section{Gas Control System}
\label{sec:gascontrol}

\begin{figure}
\begin{center}
\includegraphics[width=\linewidth]{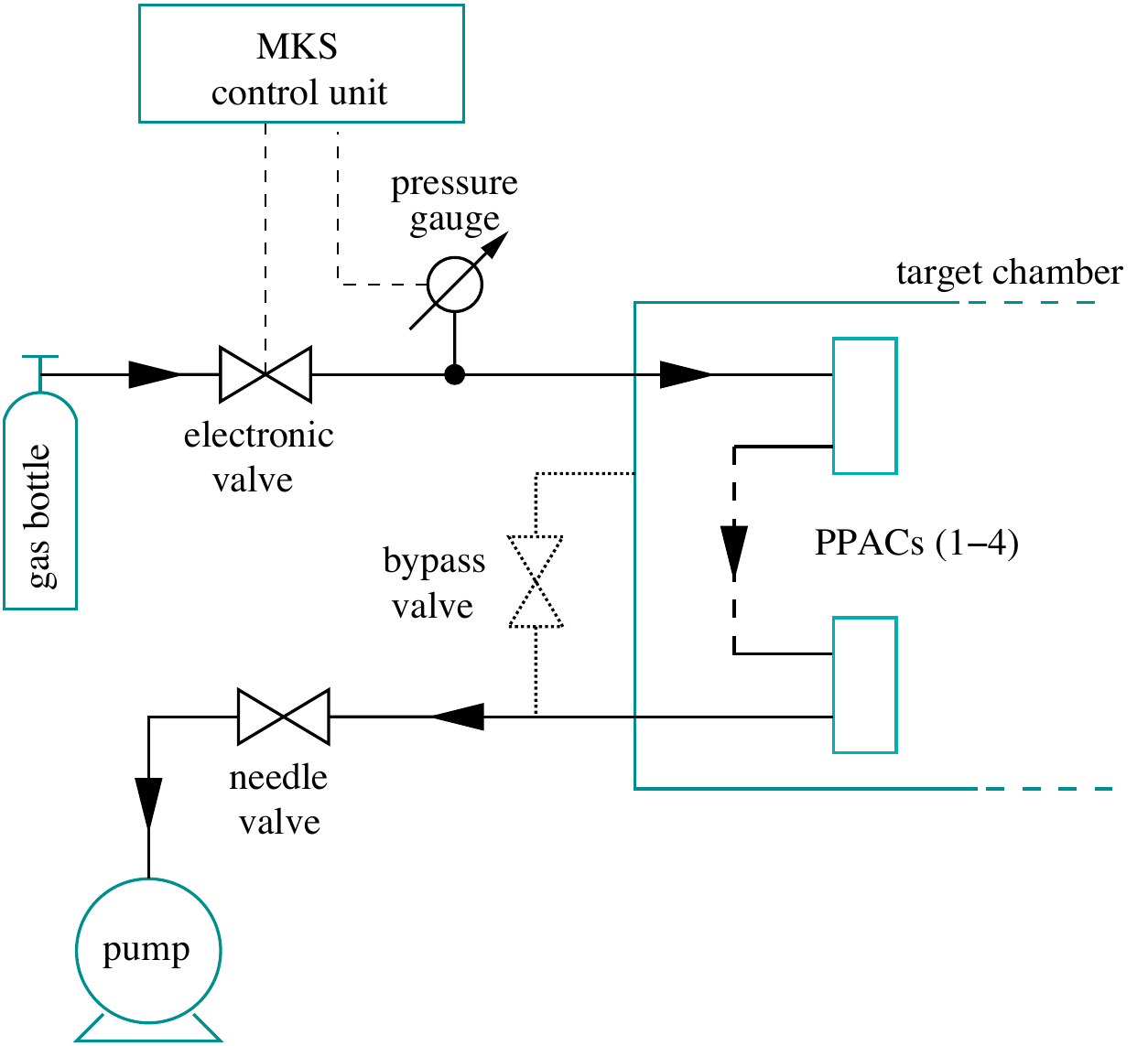}
\caption{Schematic drawing of the gas control system that keeps the pressure inside the PPAC modules constant and at the same time maintains a steady flow of isobutane gas.}
\label{fig:gas}
\end{center}
\end{figure}

During operation the avalanche process leads to polymerization of the isobutane gas, which 
affects the detection efficiency of the PPAC modules. 
It is therefore necessary to replace the gas continuously in order to maintain a high efficiency 
that is constant over time. 
The four PPAC modules are chained together with the gas flowing subsequently through all four modules. 
A continuous gas flow of $\unit[\approx 1] {ml/s}$ is necessary to maintain a good efficiency. 
At the same time the pressure inside the PPAC modules has to be kept constant to avoid fluctuations 
of the detection efficiency. 
This is achieved by the gas control system shown in Fig.~\ref{fig:gas}.

The pressure inside the detector loop is measured with an MKS 626B Baratron absolute capacitance 
manometer and maintained at a constant value by an MKS 250E gas inlet pressure controller. 
During operation the gas is continuously pumped out from the detector loop through a needle valve. 
If the pressure falls below a lower threshold the inlet controller opens an MKS 248A control 
valve until the pressure reaches an upper threshold. 
In this way the pressure is kept constant within a preset range. 
A typical pressure inside the PPAC modules is $\unit[\approx 5] {mbar}$, which is kept within 
a range of $\unit[\approx] 2{\%}$. 
The gas flow is adjusted by opening or closing the needle valve.

Particular care has to be taken when evacuating or venting the target chamber. 
The detector volume is separated from the volume of the target chamber only by the 
$\unit[1.5] {\upmu m}$ thick Mylar foil, which will easily be damaged when the pressure difference 
becomes too large. 
Tests have shown that the foils can withstand a pressure difference of $\unit[10] {mbar}$ without damage. 
A bypass valve is used to connect the detectors with the volume of the target chamber. 
In this way both volumes can be pumped down simultaneously before the detectors are filled with isobutane. 
Conversely, the gas from the detectors can be vented into the chamber vacuum through the bypass valve before both volumes are brought up to atmospheric pressure simultaneously. 

\section{Electronics and Data Acquisition}
\label{sec:electronics}

\begin{figure}
\begin{center}
\includegraphics[width=\linewidth]{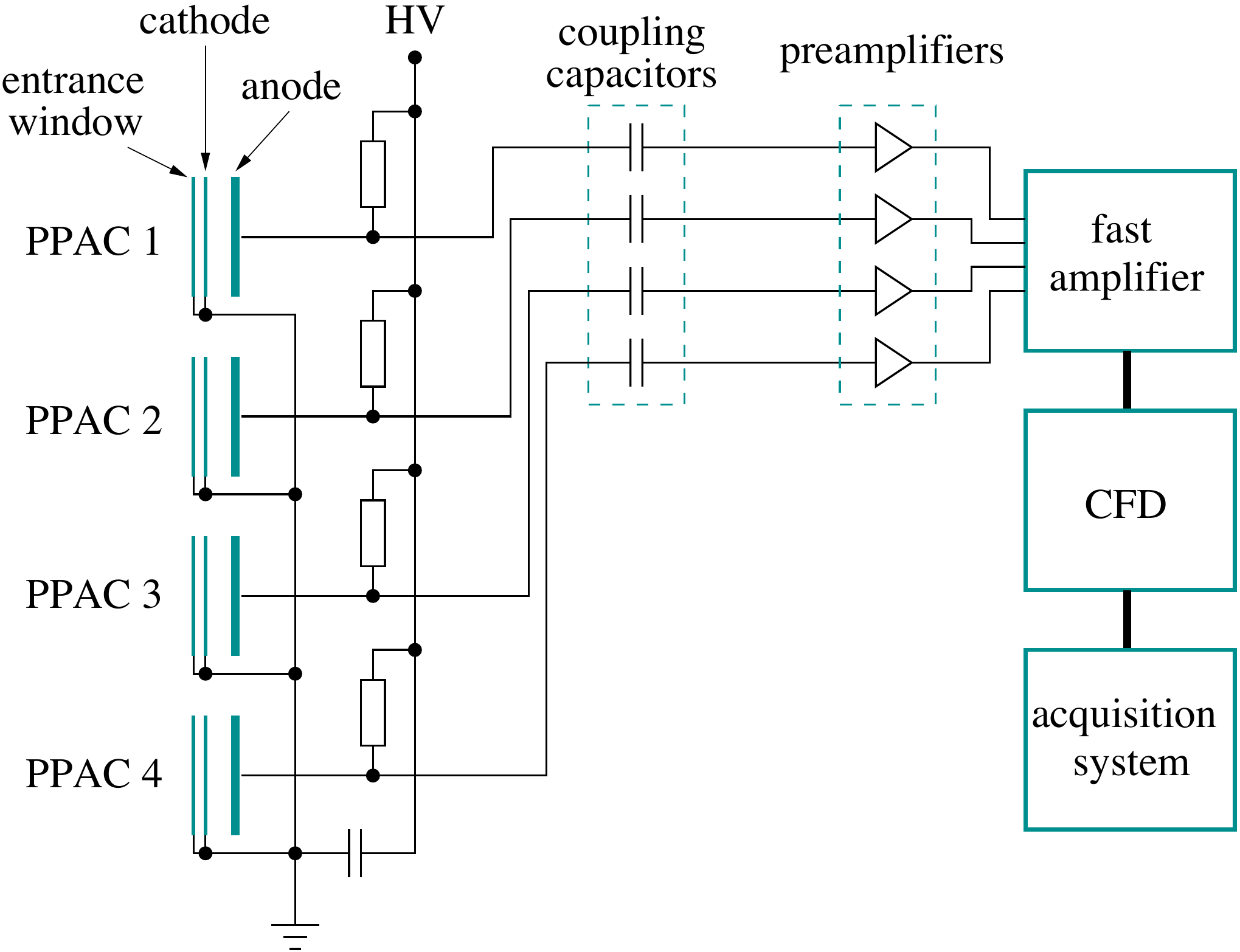}
\caption{Since anodes are fed by one high voltage supply, resistances and a coupling capacitor to the ground are needed to prevent electronic crosstalk.}
\label{fig:electr}
\end{center}
\end{figure}

A schematic drawing of the detector system and the signal processing electronics is 
shown in Fig.~\ref{fig:electr}.
To avoid the build-up of charge, both the entrance windows and cathode foils are connected 
to a common ground. 
A common positive high voltage of $\unit[\approx 400] {V}$ is applied to the anodes via 
$\unit[1] {M \Omega}$ resistors. 
The resistors, together with a $\unit[2] {nF}$ coupling capacitor to ground, prevent 
cross-talk between the four different detector modules. 
The fission fragments enter the active detector volume through the entrance window 
and cathode foil and ionize the isobutane gas along their trajectory. 
The electrostatic field accelerates the primary electrons towards the anode. 
If the voltage is high enough the primary electrons create an avalanche of secondary 
electrons in collisions with the gas. 
The avalanche effect provides sufficient charge on the anode to produce a measurable 
signal, which is taken out via $\unit[2] {nF}$ capacitors to decouple the preamplifier 
input from the high voltage. 
The detector performance was tested as a function of isobutane gas pressure in the 
detector and as a function of the applied high voltage. 
The results are described in Sect.~\ref{sec:performance}.

The anode signals pass through Ortec VT120A fast preamplifiers with a gain factor of 200, 
which are located on the outside of the target chamber in close proximity to the vacuum 
feedthroughs. 
The signals are further amplified using Tennelec TC248 shaping amplifiers. 
The fast output signals are fed into an Ortec CF8000 constant fraction discriminator. 
The logic signal of the CFD output is used in the VME-based data acquisition system 
\cite{siri} to determine the time difference between the detection of a fission fragment 
in NIFF and a light charged particle in the SiRi telescopes, which generates the event trigger. 
The pulse height of the PPAC contains no information and is not used in the data processing. 

\section{System performance}
\label{sec:performance}

The performance of the new PPAC detectors was first tested using a $^{252}$Cf 
source and its integration with the SiRi charged-particle telescopes and the CACTUS 
array of NaI scintillation detectors was tested during in-beam experiments with $^{238}$U 
and $^{237}$Np targets.

A $^{252}$Cf source of well-known activity was used to measure the efficiency of the 
individual PPAC modules and the detector as a whole. 
The count rates of the individual PPAC modules were found to be very similar to each 
other and within $\unit 4{\%}$ of the average count rate of the four modules during 
both source and in-beam measurements. 
The test with the $^{252}$Cf source showed that the PPAC modules only trigger on fission
fragments, which originate from the $\unit 3.09{\%}$ spontaneous fission branch, but 
not on the $\unit[6.1] {MeV}$ $\alpha$ particles from the predominant $\alpha$-decay branch.

\begin{figure}
\begin{center}
\includegraphics[width=\linewidth]{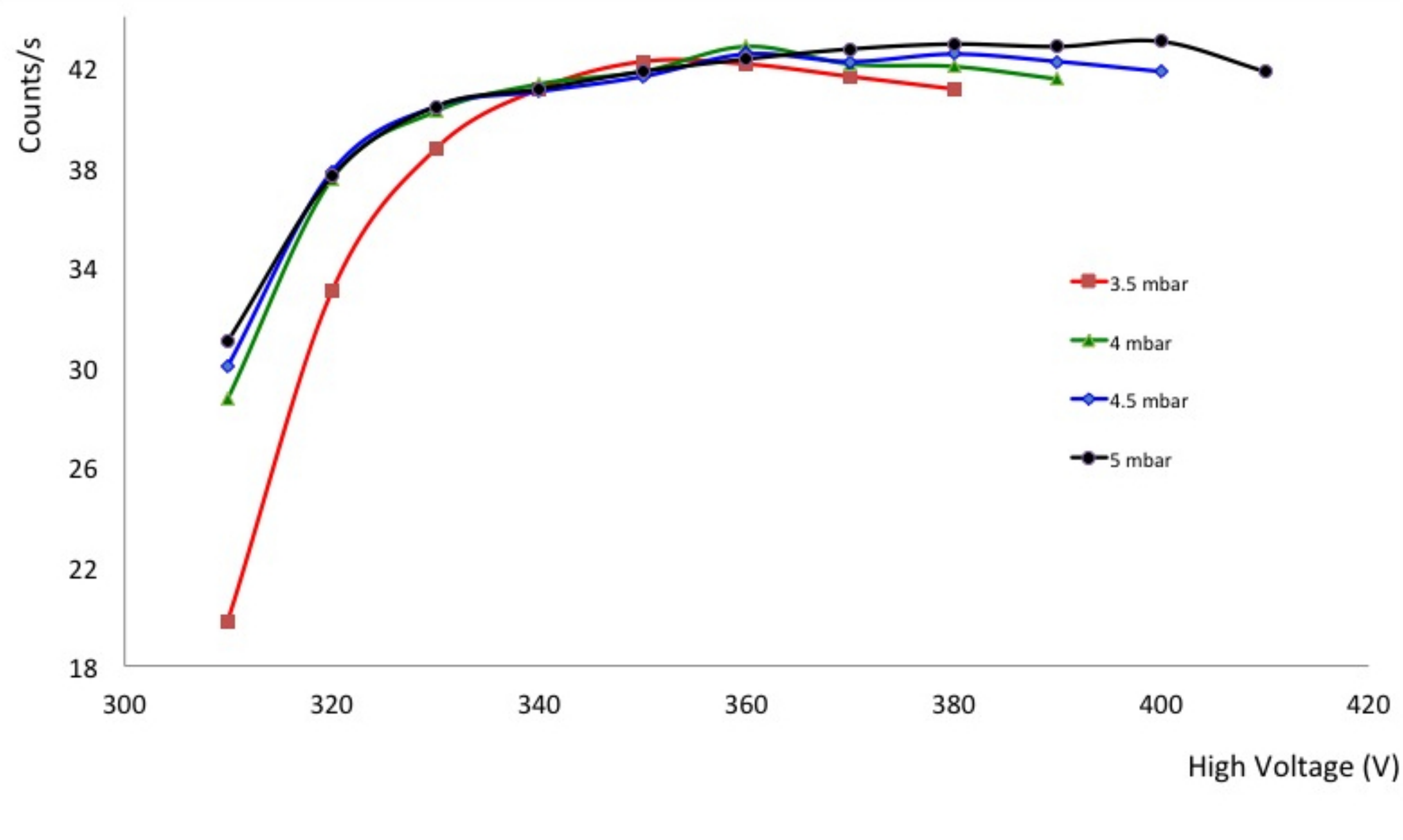}
\caption{Count rates of the total PPAC detector array as a function of high voltage for different isobutane gas pressures. A $^{252}$Cf source of well-known activity was placed at the target position of the chamber. The statistical uncertainty of the individual data points is less than 
$\unit 1{\%}$.}
\label{fig:volt_func}
\end{center}
\end{figure}

The efficiency of the total detector array to detect fission fragments was measured as a 
function of the isobutane gas pressure inside the PPAC modules and as a function of the 
applied high voltage. 
The $^{252}$Cf source was mounted at the target position for this measurement. 
The results of the measurement are shown in Fig.~\ref{fig:volt_func}. 
The statistical uncertainty of the individual measurements is better than $\unit 1{\%}$. 
The curves indicate the optimum voltage for a given gas pressure. 
Once the plateau region is reached, the efficiency increase with pressure and voltage is very small. 
To avoid damage of the thin entrance windows the detectors were only operated with gas pressures below 
$\unit[5]{mbar}$. 
The detectors are furthermore insensitive to light charged particles at such low pressure. 
Within this limit the highest efficiency for the detection of fission fragments was found for a gas 
pressure of $\unit[5]{mbar}$ and a voltage of $\unit[400]{V}$. 
These parameters were chosen for all subsequent measurements. 
Fig.~\ref{fig:volt_func} also illustrates that pressure variations of the order of $\unit 2{\%}$ 
(cf.~sect.~\ref{sec:gascontrol}) have negligible implications for the efficiency of the detectors.
Under these conditions the four PPAC modules counted fission fragments at a total rate of 
$\unit[43.2(4)] {Hz}$. 
The activity of the source was derived from a previous measurement of the $\alpha$ activity 
and, taking into account the halflife of $^{252}$Cf, an activity of  $\unit[2.54(8)] {kBq}$ 
was determined. 
With a fission branch of $\unit 3.09{\%}$ we find the fission rate to be $\unit[78.5(24)] {Hz}$. 
From the decay rate we determine the probability to detect fission events in NIFF to be $\unit 55(2){\%}$. 
We have estimated that the active areas of the four PPAC modules cover just under $\unit 60{\%}$ 
of the forward hemisphere, which corresponds to the chance that one of the two fission fragments 
enters the active volume of the detecor. 
We can therefore conclude that the intrinsic efficiency of the PPAC modules to detect an 
incoming fission fragment is well above $\unit 90{\%}$.  

\begin{figure}
\begin{center}
\includegraphics[width=\linewidth]{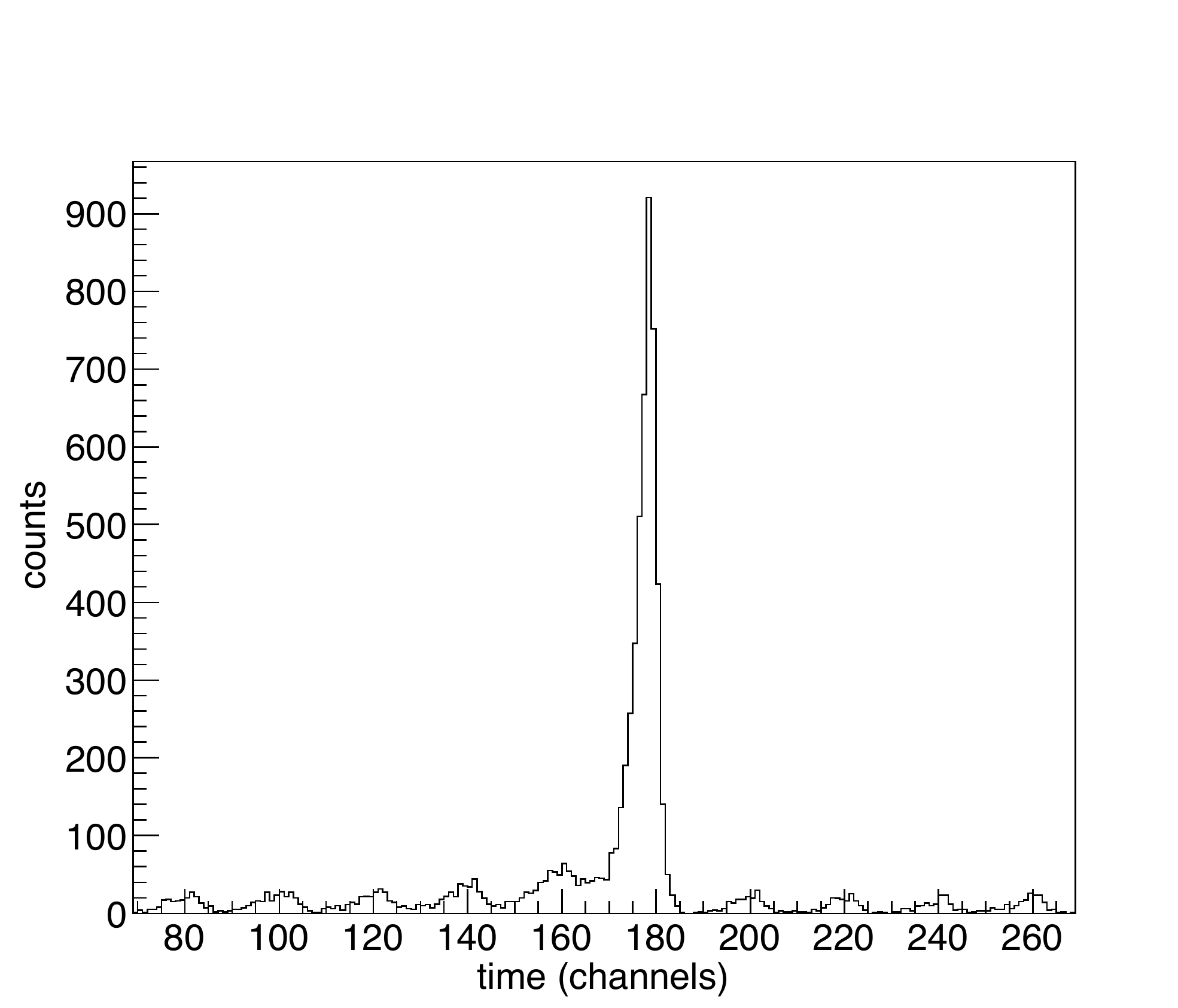}
\caption{Spectrum showing the time difference between particles detected in SiRi (start signal) and fission fragments detected in NIFF (stop signal) taken during an 
in-beam measurement with the reaction $^{238}$U(d,pf) at a beam energy of $\unit[12] {MeV}$. The smaller peaks correspond to random coincidences between different beam bursts. Each channel on the abcissa corresponds to a time interval of 
$\unit[2.4] {ns}$.}
\label{fig:timespec}
\end{center}
\end{figure}

After the extensive measurements with the $^{252}$Cf source the new detector was also 
tested during an in-beam experiment using the reaction $^{238}$U(d,pf). 
The deuteron beam energy was $\unit[12] {MeV}$ and the metallic $^{238}$U target had 
a thickness of $\unit[260] {\upmu g/cm^2}$. 
The data acquision was triggered by the logic OR of the back detectors of SiRi. 
The energy information from the thin ${\Delta}E$ front detectors and the thick back 
detectors are used to identify the charged-particle ejectiles which are detected in 
the SiRi telescopes under backward angles, essentially protons, deuterons, and tritons. 
Almost all fission fragments that are detected in the PPAC modules of NIFF are associated 
with low-energy protons detected in SiRi, {\it i.e.} the fragment originates from the fission 
of $^{239}$U above the threshold. 
The spectrum in Fig.~\ref{fig:timespec} shows the time difference between the detection of a 
proton in SiRi, which provides the start signal, and the detection of a fission fragment in 
one of the PPAC modules, which provides the stop signal. 
The peak has a width of $\unit[11]{ns}$ FWHM, which is typical for the silicon detectors of SiRi. Fast signal rise times observed for the PPAC modules suggest that the NIFF detectors are fast compared to the silicon detectors and that the time resolution is entirely dominated by the SiRi detectors."
It would in principle be possible to operate the PPAC detectors with count rates in the MHz region. 
However, in experiments where NIFF is coupled to SiRi and CACTUS the count rates are usually limited by the latter and are not expected to exceed a few kHz in NIFF.

The new PPAC detector was furthermore tested in another in-beam experiment with the 
reaction $^{237}$Np(d,pf). 
In this case the deuteron beam energy was $\unit[13.5] {MeV}$ and the target consisted 
of $^{237}$Np oxide with a thickness of $\unit[200] {\upmu g/cm^2}$ on a 
$\unit[20] {\upmu g/cm^2}$ thick carbon backing. 
The target contained a total amount of $\unit[35] {\upmu g}$ of $^{237}$Np and had an 
$\alpha$ activity of $\unit[0.9] {kBq}$. 
As was the case for the source measurement with $^{252}$Cf, the $\alpha$ particles from 
the decay of $^{237}$Np do not ionize the isobutane gas sufficiently for the PPAC 
detectors to give any signals. 
The decay $\alpha$ particles are furthermore stopped in the ${\Delta}E$ front detectors 
of SiRi. 
Without reaching the back detectors of the telescopes they do not trigger the data 
acquisition, and all recorded events are due to deuteron-induced reactions on the target.

\begin{figure}
\begin{center}
\includegraphics[width=\linewidth]{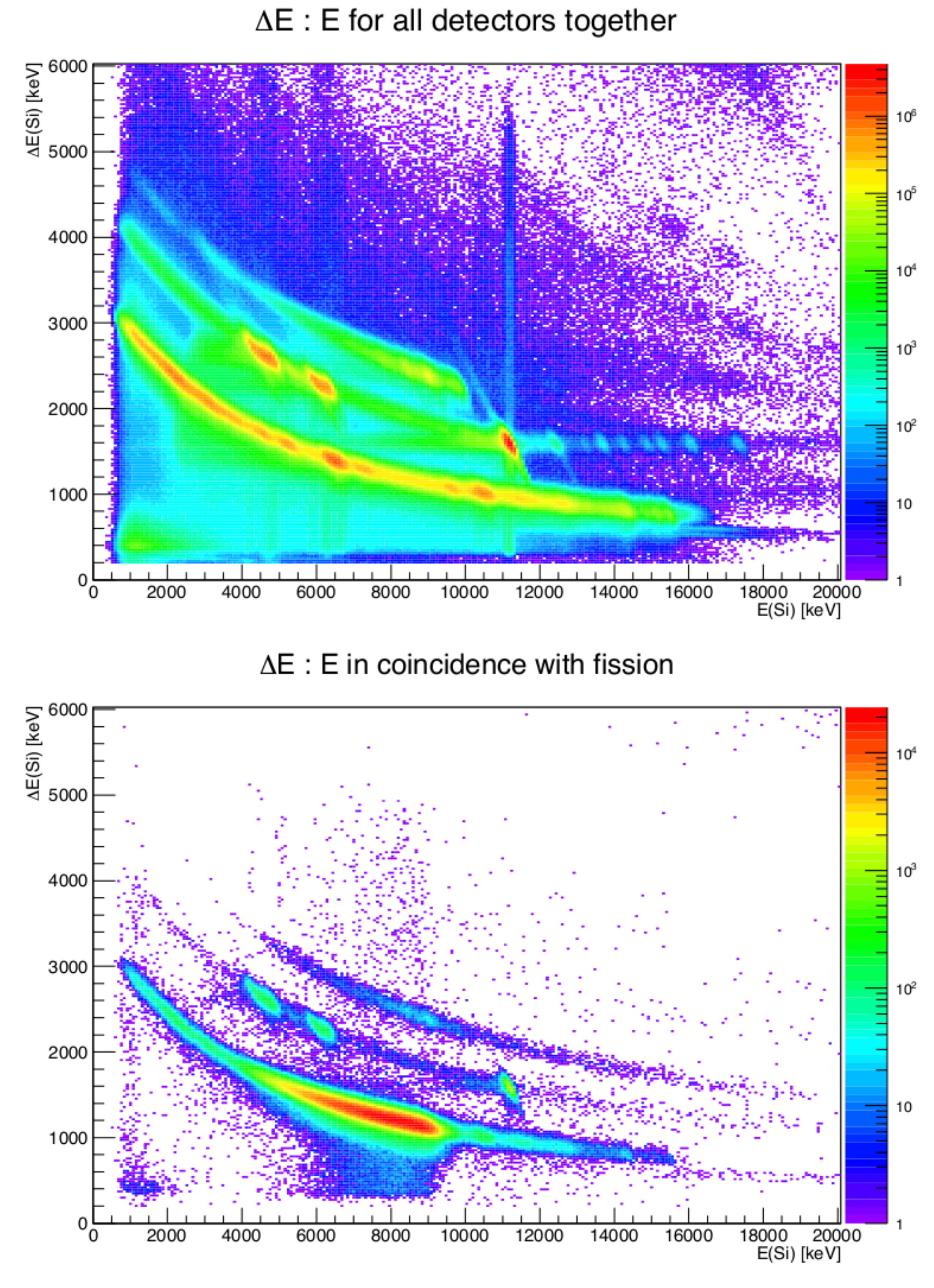}
\caption{Particle identification plots showing the energy loss ${\Delta}E$ in the front detectors of SiRi versus the the total energy $E$ of the particles ejected following the 
d+$^{237}$Np reaction. The upper spectrum shows all recorded events, the one on the bottom only those that were recorded in coincidence with a fission fragment in NIFF. The three separate loci visible in the total identification plot correspond to (bottom to top) protons, deuterons, and tritons.
Fission events are only associated with protons below a certain energy, which corresponds to fission of $^{238}$Np above the fission threshold.}
\label{fig:E_dE}
\end{center}
\end{figure}

The spectra of Fig.~\ref{fig:E_dE} show ${\Delta}E-E$ particle identification plots 
from the d+$^{237}$Np experiment. 
The upper spectrum contains all events recorded in SiRi without any 
condition on the NIFF fission fragment detector. 
Three distinct curves are observed that are associated with (bottom to top) protons, 
deuterons, and tritons. 
The strongest peak at the highest deuteron energy is due to elastic scattering of 
the projectiles on $^{237}$Np. 
The other strong peaks associated with detected deuterons are due to elastic scattering 
on $^{16}$O and $^{12}$C, respectively. 
The strong peaks associated with detected protons are due to inelastic scattering to 
excited states in $^{16}$O and $^{12}$C. 
The lower spectrum shows the same particle-identification plot with 
the condition that a fission fragment was detected in NIFF. 
As can be seen, fission events are associated with the detection of protons of a certain 
energy range. 
The few events where fission was detected together with deuterons (note the logarithmic scale) 
are due to random coincidences. 
Reactions in which protons are emitted with the maximum energy of $\unit[{\sim}16] {MeV}$ leave $^{238}$Np 
in the ground state. 
The lower the energy of the ejectile the higher is the excitation energy of the reaction product. 
Fission is observed for proton energies lower than $\unit[{\sim}10] {MeV}$, which corresponds 
to an excitation energy of $\unit[{\sim}6] {MeV}$ and coincides with the fission threshold in $^{238}$Np. 
The example illustrates how the new NIFF detector can be used to determine fission cross sections 
and the shape of the fission barrier for more exotic, less-known actinide isotopes.

\begin{figure}
\begin{center}
\includegraphics[width=\linewidth]{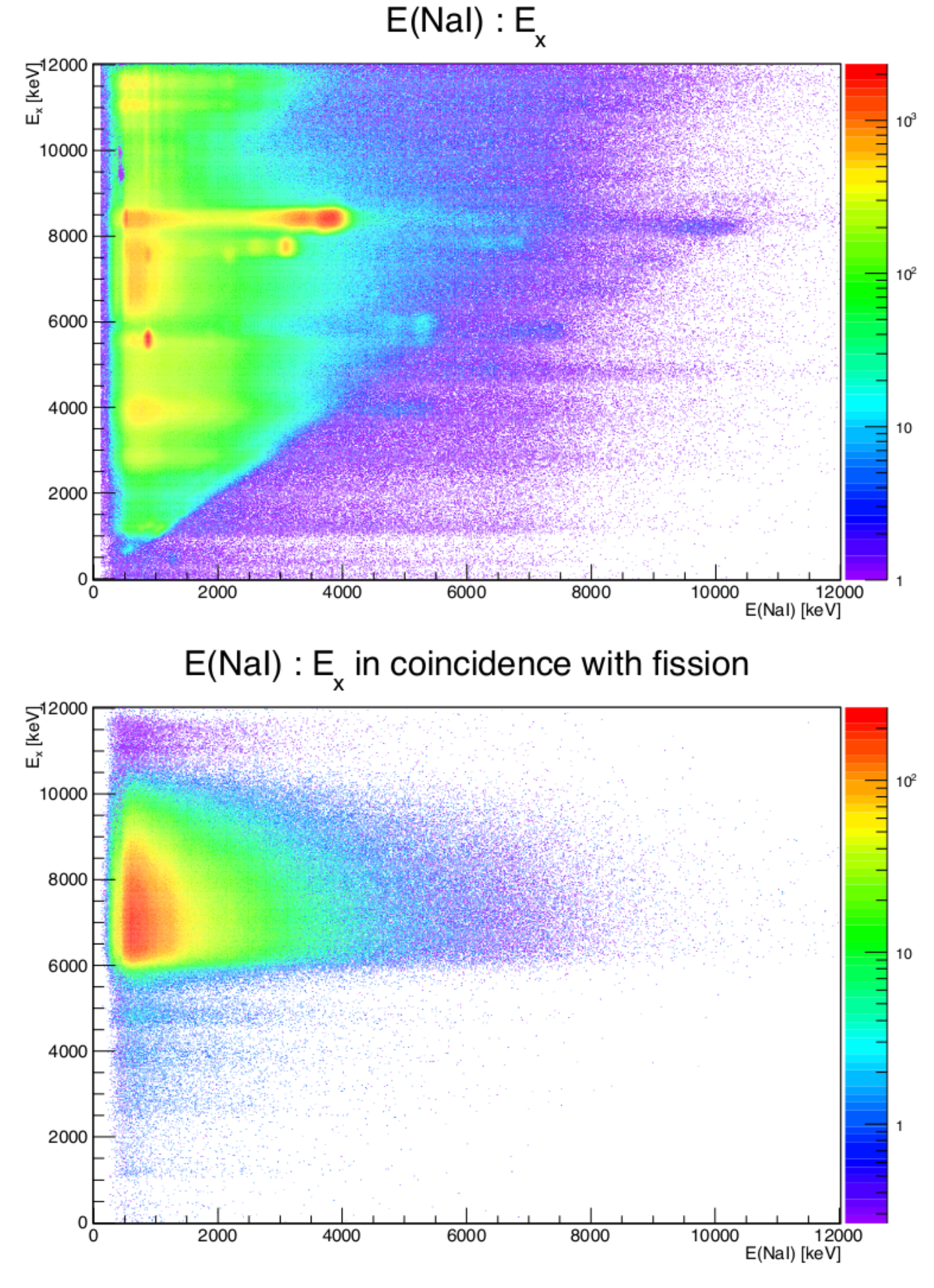}
\caption{Excitation energy versus $\gamma$-energy matrices for the reaction $^{237}$Np(d,p)$^{238}$Np. The excitation energy is obtained from the measured kinetic energy of the protons and the reaction kinematics. The spectrum on the bottom was obtained with the additional condition that a fission fragment was detected in the NIFF detector in coincidence with a proton in SiRi and $\gamma$-rays in CACTUS.}
\label{fig:m_alfna}
\end{center}
\end{figure}

The detection of fission fragments in NIFF allows identifying $\gamma$-rays that are 
associated with the fission process. 
Fig.~\ref{fig:m_alfna} shows $\gamma$-ray spectra as a function of excitation energy 
for $^{238}$Np. 
In a first step proton events were selected in the ${\Delta}E-E$ identification matrix. 
Next a time gate was applied to select $\gamma$-rays that were recorded in CACTUS in prompt 
coincidence with the protons. 
The excitation energy of the nucleus can be determined from the measured proton energy and 
the reaction kinematics. 
In this way it is possible to extract the $\gamma$-ray spectrum for a given bin of excitation 
energy in $^{238}$Np \cite{OsloMethod}. 
The resulting two-dimensional spectrum is shown in Fig.~\ref{fig:m_alfna} on the top. 
A few discrete $\gamma$-ray transitions are clearly visible. 
These originate from excited states in $^{13}$C and $^{17}$O, which are populated in (d,p) 
reactions on the carbon backing and oxygen in the target, and can be easily subtracted. 
The lower part of Fig.~\ref{fig:m_alfna} shows the same data set with the additional condition that a fission fragment was detected in NIFF. 
From these spectra it is possible to extract information on the competition between 
$\gamma$ decay and fission as a function of excitation energy. Under the assumption that 
the formation and decay of a compound nucleus are independent of each other, such measurements 
can be used to determine cross sections for compound nuclear reactions via the so-called surrogate method. 
The technique is particularly useful in cases where the direct measurement of (n,$\gamma$) 
and (n,f) cross sections is not feasible because the required actinide targets are too short-lived.

\section{Conclusion}
\label{sec:conclusion}

A new fission fragment detector based on parallel-plate avalanche counters was developed 
at the University of Oslo. 
It was designed to be used inside the CACTUS array of large-volume NaI scintillation 
detectors together with the SiRi charged-particle telescope array. 
The new detector consists of four PPAC modules which cover close to $\unit 60{\%}$ of 
the forward hemisphere. 
The efficiency of the detector was measured with a $^{252}$Cf source. 
The intrinsic efficiency for heavy ions was found to be well above $\unit 90{\%}$, resulting 
in an efficiency for the detection of one of the two fragments from a fission event of $\unit 55(2){\%}$. 
The PPAC detectors are insensitive to light ions, electrons, neutrons, or photons. 
The integration of the fission detector into the existing CACTUS and SiRi data acquisition 
system was tested during in-beam experiments with deuteron beams and $^{238}$U and $^{237}$Np targets. 
The detector has excellent time resolution and allows measuring particle-$\gamma$-fission coincidences. 
The combination of SiRi, CACTUS, and NIFF is a powerful setup to investigate the competition of 
$\gamma$ decay and fission in highly excited actinide nuclei. 
By employing charged-particle induced surrogate reactions, such measurements can provide valuable 
information on neutron-induced reaction cross sections which are not accessible by direct measurements.

\section*{Acknowledgments}

Financial support from the University of Oslo for the construction of the detector is greatfully acknowledged. 
We would also like to thank the staff at the instrumental workshop of the Department of Physics for their help in 
the manufacture of the parts for the detector and the operators of the Oslo Cyclotron Laboratory for providing the beam.
The work has also been supported by the  Hungarian OTKA Foundation No. K 106035 and by the National Excellence 
Program T\'AMOP 4.2.4.A/2-11-1-2012-0001.

\end{document}